\documentclass[sigconf]{acmart}

\AtBeginDocument{%
  }

\setcopyright{acmlicensed}
\copyrightyear{2024}
\acmYear{2024}
\acmDOI{XXXXXXX.XXXXXXX}

\acmConference[UbiComp '24]{2024 UbiComp Mental Health Sensing and Intervention workshop}{October 6, 2024}{Melbourne, Australia}

\usepackage{tcolorbox}
\usepackage{makecell}
\usepackage{xcolor}
\usepackage{soul}
\usepackage{graphicx}
\usepackage{subfigure}
\usepackage{cleveref}
\usepackage{adjustbox}
\usepackage{graphicx}
\usepackage{graphicx}
\usepackage{booktabs}
\usepackage{multirow}
\usepackage{caption}
\usepackage{amsmath}

\definecolor{mygreen}{RGB}{0,128,0}
\definecolor{myred}{RGB}{255,0,0}
\begin{document}

\title{Leveraging LLMs to Predict Affective States via Smartphone Sensor Features}



\author{Tianyi Zhang, Songyan Teng, Hong Jia, Simon D'Alfonso}
\email{{t.zhang59, stteng}@student.unimelb.edu.au,{hong.jia, dalfonso}@unimelb.edu.au}

\affiliation{%
  \institution{University of Melbourne}
  \country{Australia}
}





\renewcommand{\shortauthors}{Zhang et al.}

\begin{abstract}
As mental health issues for young adults present a pressing public health concern, daily digital mood monitoring for early detection has become an important prospect. An active research area, digital phenotyping, involves collecting and analysing data from personal digital devices such as smartphones (usage and sensors) and wearables to infer behaviours and mental health. Whilst this data is standardly analysed using statistical and machine learning approaches, the emergence of large language models (LLMs) offers a new approach to make sense of smartphone sensing data. Despite their effectiveness across various domains, LLMs remain relatively unexplored in digital mental health, particularly in integrating mobile sensor data. Our study aims to bridge this gap by employing LLMs to predict affect outcomes based on smartphone sensing data from university students. We demonstrate the efficacy of zero-shot and few-shot embedding LLMs in inferring general wellbeing. Our findings reveal that LLMs can make promising predictions of affect measures using solely smartphone sensing data. This research sheds light on the potential of LLMs for affective state prediction, emphasizing the intricate link between smartphone behavioral patterns and affective states. To our knowledge, this is the first work to leverage LLMs for affective state prediction and digital phenotyping tasks.
\end{abstract}





\received{18 June 2024}

\maketitle

\section{Introduction}

With the increasing prevalence of mental health issues among young adults, particularly within the student population, there is significant rationale for researching and developing systems to monitor mental health, including affective states. Daily mood monitoring may aid early detection of potential mental health issues in timely intervention and support \citep{beames2021prevention}. Digital phenotyping, also called personal sensing, may serve as one of the potential solutions. Utilizing digital devices to collect behavioral traces seamlessly to analyse and predict mental health status, digital phenotyping is a potentially effective tool to understand the relationship between human activities and mental health or wellbeing.

Smartphones are a premium device for digital phenotyping due to their ubiquity and diverse data collection capacities, including collection of multi-sensor data (e.g. GPS and accelerator) and also captures of virtual social behaviors, software usage, and screen interactions, thereby capturing the intricate relationship between humans and technology. To deliver emotional wellbeing predictions, powerful classic machine learning models such as deep neural networks \citep{lecun2015deep}, gradient boosting \citep{chen2016xgboost}, and transformers \citep{vaswani2017attention} often face overfitting issues and are not suitable for small datasets typically found in mental healthcare. Therefore, innovative and effective new approaches are required to assist and improve digital phenotyping techniques and methods.

Recently, large language models (LLMs) have emerged as powerful tools for natural language generation, natural language understanding, and contextual comprehension. Their ability to leverage pre-trained knowledge makes them applicable to various research problems across different domains, demonstrating superior performance in tasks such as sentiment analysis \citep{vaswani2017attention}, activity recognition \citep{liu2023large}, and information retrieval \citep{zhai2001model}. The inherent nature of foundational LLMs, trained on vast online corpora that document human behavior and psychological knowledge, underscores their potential to reveal more nuanced healthcare information compared to traditional machine learning models in personal sensing, as exemplified by recent studies \citep{le2021machine}. While these initial studies provide a direction for utilizing LLMs in addressing general healthcare issues, the application of LLMs to digital phenotyping for mental health (DPMH) and emotional wellbeing has been scarcely studied, particularly regarding the integration of smartphone sensing data.

In contrast to broader data-driven tasks, the intricacies of DPMH and emotional wellbeing involve smaller-scale personal and sensitive data and subjective measures. The nature of this sensitivity and subjectivity renders the DPMH tasks more challenging because of contextual dependence and difficulty in identifying generalized patterns. Additionally, in the nascent field of DPMH, the exploration of multimodal or multisensor activities in daily behavioral data collected via smartphones remains limited, despite its potential to indicate individuals' psychological fluctuations. Given that LLMs are well-known for capturing contextual information, they are a natural choice for analyzing digital traces inferred from smartphones. Compared with traditional ML models focused on analysing numeric features produced from individual sensors, LLMs can analyse generated lifecycle descriptions from data collected across multiple smartphone sensors at a higher level of abstraction. Then, powered by prior knowledge, LLMs are capable of unveiling the latent associations between behavioral patterns and individuals' mental health or affective status. As such, our research aims to explore whether LLMs can effectively predict affective states with minimal amounts of data, utilizing human behaviors and interactions with smartphones as input data.

In this study, we investigate the relationship between behavioral features collected from smartphone sensors and the affects of university students. We also demonstrate the capability of zero-shot and few-shot embedding LLMs to infer affective states based on smartphone-captured human activities. Our results suggest a discernible connection between smartphone-sensed activities of university students and their affective states, which LLMs interpret through their chains of thought. To the best of our knowledge, this is the first paper to delve into grounding LLMs with mobile sensing features for affect prediction tasks.



\section{Method and Experiments}
In this section, we describe the study design (\autoref{section_method}), data collection (\autoref{section_method}), data engineering (\autoref{section_data_processing}) and tasks delivered to LLMs (\autoref{section_task_description}). Specifically, we detail in \S\ref{section_task_description} the prompts for LLMs on multiple emotional state prediction tasks, including zero-shot prompting, few-shot prompting and chain-of-thought reasoning.

\vspace{-1em}
\subsection{Participants and Data Collection} \label{section_method}

\textbf{Study Design.}To explore the capability of LLMs in predicting individuals' general feelings based on smartphone-collected passive data, we investigated a subset of data obtained from a digital phenotyping study of Australian university students conducted in 2023. In this study \citep{dalphonso2024}, approximately 150 university students were observed over a full semester (17 weeks). Data was gathered from multiple smartphone sensors and a variety of psychometric assessments were conducted. Of the approximately 150 participants in this study, we chose to analyse the data of 10 students for this investigation. The study utilised the AWARE-Light smartphone sensing app \citep{van2023aware} to collect passive sensing data. Weekly self-reported assessments were collected using an emailed link to a Qualtrics questionnaire. Our study continuously collected sensor data for 17 weeks from the primary smartphones university students used. Informed by our own previous work as well as existing literature, for example \cite{politou2017survey}, the following sensors were chosen: battery levels, screen unlocks, location information, application usage, keyboard usage, and communication traces (calls and messages). Data collection was in accordance with ethics approval from the University of Melbourne.

~\\\textbf{Self-reported Measures.} At the end of each week, participants were prompted to complete the International version of the Positive and Negative Affect Schedule (I-PANAS-SF) questionnaire \citep{thompson2007development}. This questionnaire comprises 10 items rated on a Likert scale ranging from 1 (indicating "Never") to 5 (indicating "Always"). Five of the items indicate positive affects (active, determined, attentive, inspired, alert) and the other five indicate negative affects (upset, hostile, ashamed, nervous, afraid).

\subsection{Data processing} \label{section_data_processing}
We first cleaned the raw data from the self-reported questionnaire and passively collected smartphone data to remove duplicates and corrupted entries before carrying out data engineering. We utilized the RAPIDS tool \citep{vega2021reproducible} to generate and select 77 behavioral features at the daily level of granularity. These features include metrics such as time spent at home, the number of received messages, and duration of screen unlocks per day. Missing passive data was treated as such when constructing prompts, without any imputation.

~\\Previous studies focusing on machine learning models for predicting individuals' mental health scales have typically targeted the total score of a questionnaire, such as the PHQ-9 for depressions or GAD-7 for anxiety. In terms of I-PANAS-SF, a total score is not feasible, as the questionnaire is not designed to provide an overall score. Whilst it is possible to sum up the five positive items to get a total positive score, and likewise for negative items, our approach was to construct a prediction for each individual scale item. By doing so, we acknowledge that each item represents a distinct affect, making our predictions more comprehensive. For instance, predicting solely the total score of I-PANAS-SF negative affect provides little insight into the extent to which a person felt upset or nervous. Therefore, we constructed the task in a manner where LLMs are prompted to predict scores for each single item of the I-PANAS-SF separately within one inference.

~\\Given the nature of LLMs as language models, and their known limitations in handling numerical raw data \citep{spathis2023first}, we opted to reference participants' daily features with concise English descriptions. Each feature description is presented sequentially, preceded by the corresponding date and time range in the format \textit{YYYY-MM-DD HH:MM:SS to YYYY-MM-DD HH:MM:SS} (e.g., \textit{2023-08-02 00:00:00} to \textit{2023-08-02 23:59:59}). We intentionally kept the time series information for training purposes to understand if the LLM has the capacity to complete the tasks rather than build a robust and usable predictive model.

\subsection{Task description} \label{section_task_description}
To investigate whether LLMs can grasp the connection between smartphone-derived behaviours/contexts and affective states, we conducted zero-shot and few-shot tasks on I-PANAS-SF. Each week's description of daily activities consists of approximately 5,000 tokens, and we restricted our experimentation to Gemini 1.5 Pro \citep{reid2024gemini} due to resource limitations in other LLMs. The LLM was tasked with providing Likert scores rather than textual descriptions. To ensure deterministic responses from the LLMs, we adjusted the temperature to zero. Although Gemini still yields slightly different answers each time, the predictions remain relatively consistent across low (i.e., 1 or 2) and high (i.e., 4 or 5) scores. 

In this exploratory study, we assigned individual tasks to 10 subjects and randomly divided the 17-week data into a 10-week training dataset and a 7-week test dataset. We performed a repeated random train-test split on the 10-participant dataset three times. In the few-shot prompting method, we incrementally added one more example from the training dataset for each shot, starting with one shot and continuing until we reached ten shots, to test the individual instances in the test dataset.

\textbf{Zero-Shot.} We conducted zero-shot as the baseline of our study for comparison. The prompt was constructed following Figure \ref{zero_shot_prompt}:

\begin{figure}[htbp]
\centering
\fbox{%
    \parbox{1\columnwidth}{
    \fontsize{6}{7}\selectfont
    Below is a description of a university student's activities over a week, gathered from their smartphone sensors. Based on the descriptions provided, select the option that best represents how the student felt for the provided week's description below for the following feelings: \\

1. Active \\
\textit{...Determined, Attentive, Inspired, Alert, Upset, Hostile, Ashamedm, Nervous...}\\
10. Afraid \\

For each feeling, choose a Likert score ranging from 1 to 5 that best represents how the student generally felt during the week where 1 represents Never and 5 represents Always.  \\
\\
Description of the student's activities for the future week: \textit{\{feature description\}} \\
\\
Provide your choices in the following form with no other reasoning:\\
\\
Active: [predicted number]\\
\textit{...Determined, Attentive, Inspired, Alert, Upset, Hostile, Ashamedm, Nervous...}\\
Afraid: [predicted number]
    }
}
\caption{Prompt for Zero-Shot Tasks}
\label{zero_shot_prompt}
\end{figure}

\noindent\textbf{Few-Shots.} For each few-shot task, we randomly selected one or more weeks of data from the training set as labeled data. The prompts for the few-shot tasks were constructed following Figure \ref{few_shot_prompt}:

\begin{figure}[htbp]
\centering
\fbox{%
    \parbox{1\columnwidth}{
    \fontsize{6}{6.8}\selectfont
Given a series of descriptions detailing a university student's weekly activities collected from their smartphone sensors, along with their corresponding feelings, your task is to identify patterns between the student's activities and feelings. Based on these patterns, make predictions for the student's feelings according to their future activities.\\

\begin{center}
\noindent\fbox{%
    \parbox{0.95\columnwidth}{
    \fontsize{6}{7}\selectfont
According to the following behaviors of the student during a week, how active they felt is \textit{{score}}, how determined they felt is \textit{{score}}, how attentive they felt is \textit{{score}}, how inspired they felt is \textit{{score}}, how alert they felt is \textit{{score}}, how upset they felt is {score}, how hostile they felt is \textit{{score}}, how ashamed they felt is \textit{{score}}, how nervous they felt is \textit{{score}}, how afraid they felt is \textit{{score}}: \textit{{weekly description}}
}}

\end{center}
    {\raggedleft \textit{* <number-of-shot> weeks} \\}
    {\raggedright }
    
    Based on the patterns you learnt from the data provided, select the option that best represents how the student felt for the future week's description below for the following feelings:\\
\\
\textit{<10 listed I-PANAS-SF items>}\\
\\
For each feeling, choose an Likert score ranging from 1 to 5 that best represents how the student generally felt during the week where 1 represents Never and 5 represents Always. \\
\\
Description of the student's activities for the future week: \textit{\{feature description\}} \\
\\
Provide your choices in the following form with no other reasoning:\\
\\
\textit{<10 listed I-PANAS-SF items>: [predict number]}
    }
}
\caption{Prompt for Few-Shot Tasks}
\label{few_shot_prompt}
\end{figure}

\textbf{Chain-of-Thought.} We constructed chain-of-thought tasks to better understand how the LLM produced its predictions and evaluate its logical inference ability. To do this, we modified the prompt by changing the sentence ``\textit{Provide your choices in the following form with no other reasoning}" to ``\textit{Provide your choices in the following form with reasoning for each item.  The reasoning should be based on the comparison of provided student's weekly behaviors: {item}: [predicted number and reasoning].}"

\section{Results \& Discussion}

In this section, we first introduce the metrics used to evaluate the LLM's predictions. Next, we discuss the zero-shot and few-shot performance of these models. Finally, we present a chain-of-thought analysis of the LLM with a zero-shot case study.

\textbf{Metrics.} To assess the accuracy of our predictions, we employed a macro calculation approach for both the Mean Absolute Error (MAE) and the relative error (\(\epsilon\)). This approach involved two main steps: first calculating the error metrics for each participant individually, and then averaging these metrics across all participants.

The overall Mean Absolute Error (MAE) is calculated as follows:
\[
\text{MAE}_{\text{overall}} = \frac{1}{N} \sum_{i=1}^{N} \left( \frac{1}{n_i} \sum_{j=1}^{n_i} \left| P_{ij} - T_{ij} \right| \right)
\]
where \( P_{ij} \) represents the predicted value, \( T_{ij} \) represents the true value for the \( j \)-th observation of the \( i \)-th participant, \( n_i \) is the number of observations for participant \( i \), and \( N \) is the total number of participants.

~\\The mean of the true values for each participant is calculated as:
\[
\bar{T}_i = \frac{1}{n_i} \sum_{j=1}^{n_i} T_{ij}
\]

Finally, the overall relative error is obtained by averaging the relative errors of all participants:
\[
\epsilon_{\text{overall}} = \frac{1}{N} \sum_{i=1}^{N} \left( \frac{\text{MAE}_i}{\bar{T}_i} \times 100\% \right)
\]



\textbf{Zero-shot performance.} Table \ref{single_items} illustrates that zero-shot Mean Absolute Errors (MAEs) indicate relatively underperformed predictions, averaging 1.65 out of 5 ($\epsilon = 40.9\%$). The zero-shot MAE values across all ten participants exhibit minimal variability or dispersion ($\text{std}{\text{total}}$ = 0.01, $\text{std}{\text{pos}}$ = 0.06, $\text{std}_{\text{neg}}$ = 0.02), indicating a consistent predictive ability of the LLMs across various subjects.

The LLM performs similarly for both positive ($MAE = 1.62\%$, $\epsilon = 38.1\%$) and negative ($MAE = 1.67\%$, $\epsilon = 43.8\%$) affects but demonstrates varying levels of performance across different items. Specifically, it excels with the "Alert" item ($MAE = 0.87\%$, $\epsilon = 32.9\%$) in zero-shot scenarios while demonstrating poorest performance with "Afraid" ($MAE = 2.51\%$, $\epsilon = 48.7\%$) and "Nervous" ($MAE = 2.29\%$, $\epsilon = 54.3\%$). These differences may stem from varying levels of prior knowledge within the LLM or distinct indications of daily behaviors collected via smartphones across different affective states. In short, our results suggest that while LLMs possess a relatively limited capability to infer personal affective states in zero-shot, their performance remains consistent across different subjects based on observations of daily behavioral features.

\textbf{Few-shot learners.} To explore if LLMs are effective few-shot learners for the association between human activities and affects, few-shot tasks were designed to showcase LLMs' predictive power for affective states. A consistent trend is revealed that as more shots are provided to the LLM, the MAE decreases for each I-PANAS-SF item, as shown in Table \ref{single_items}. In particular, the one-shot approach emerges as the most effective, with performance exhibiting slight fluctuations thereafter. Specifically, the MAE of nine I-PANAS-SF items decreases in one-shot learning, with a mitigation of an average 10.0\% for positive affects and 43.5\% for negative affects. As more examples are provided, the rate of performance improvement decreases. 

Meanwhile, when averaging the MAE for positive and negative affects respectively, the LLM performs better for negative affects compared to positive affects in one-shot learning, while they perform similarly in ten-shot learning ($MAE_{pos} = 0.75$, $MAE_{neg} = 0.74$). This suggests that the LLM attains a similar level of knowledge more rapidly for I-PANAS-SF negative affects. 

When comparing individual items of the I-PANAS-SF, the LLM performs optimally with "Hostile" ($MAE_{average} = 0.66$) and least effectively with "Active"($MAE_{average} = 1.18$) and "Nervous" ($MAE_{average} = 1.15$) overall. Interestingly, it demonstrates the most significant improvement on "Afraid" with additional examples with a 69.7\% decrease in MAE between zero-shot and ten-shot. Note that although the performance across the ten items varies, there is a converging trend among the results when providing more examples to the LLMs, as depicted by the change in standard deviation of MAE from 0.55 in zero shot to 0.09 in ten shots. This suggests that provided with more examples, the performance gap across individual I-PANAS-SF items depending solely on the LLM's prior knowledge tend to diminish due to the LLM's enhanced information about the subject.

\begin{figure*}[ht]
  \centering
  \subfigure[]{\includegraphics[trim=0pt 20pt 0pt 0pt, clip,width=0.24\textwidth]{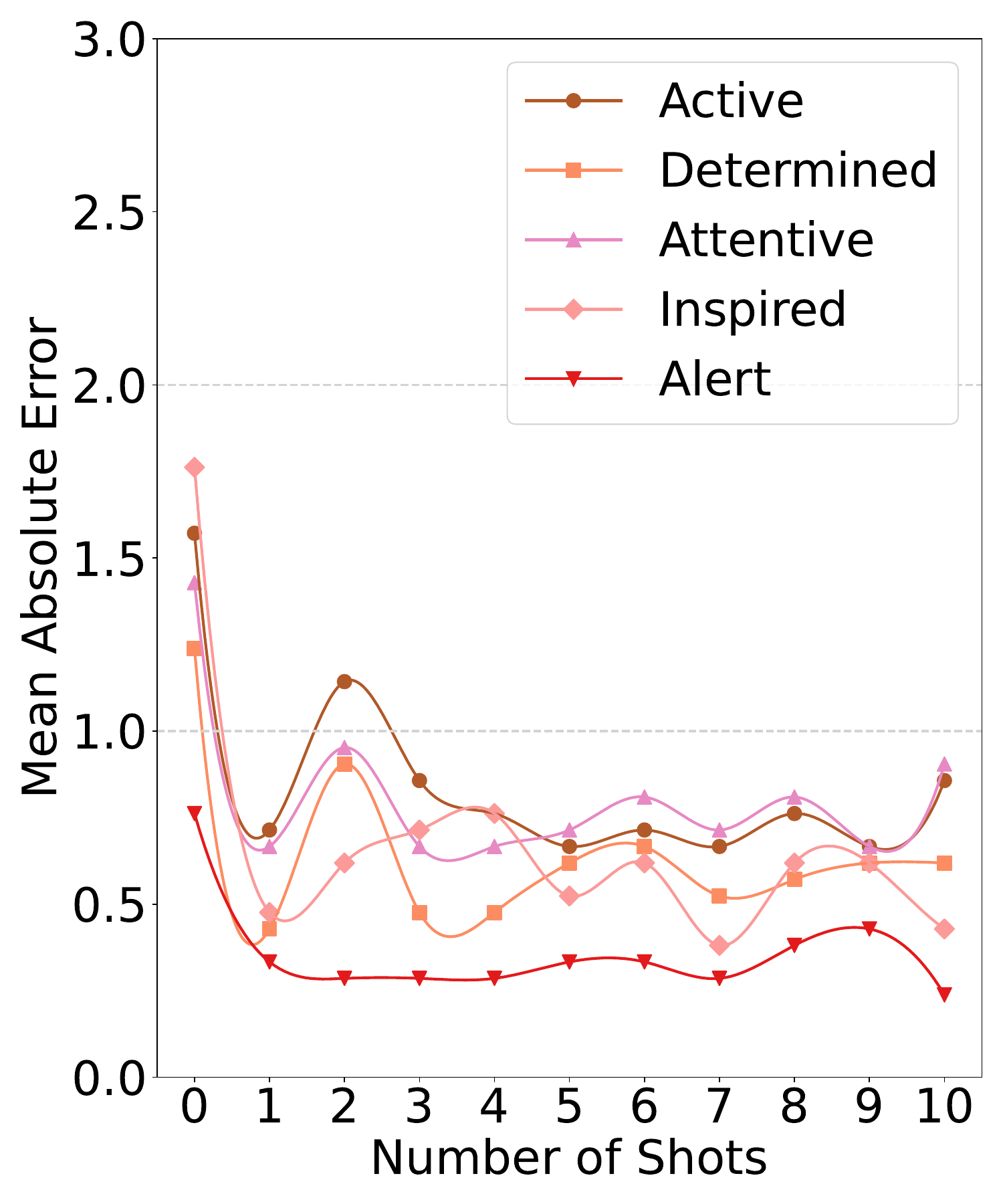}} 
  \subfigure[]{\includegraphics[trim=0pt 20pt 0pt 0pt, clip,width=0.24\textwidth]{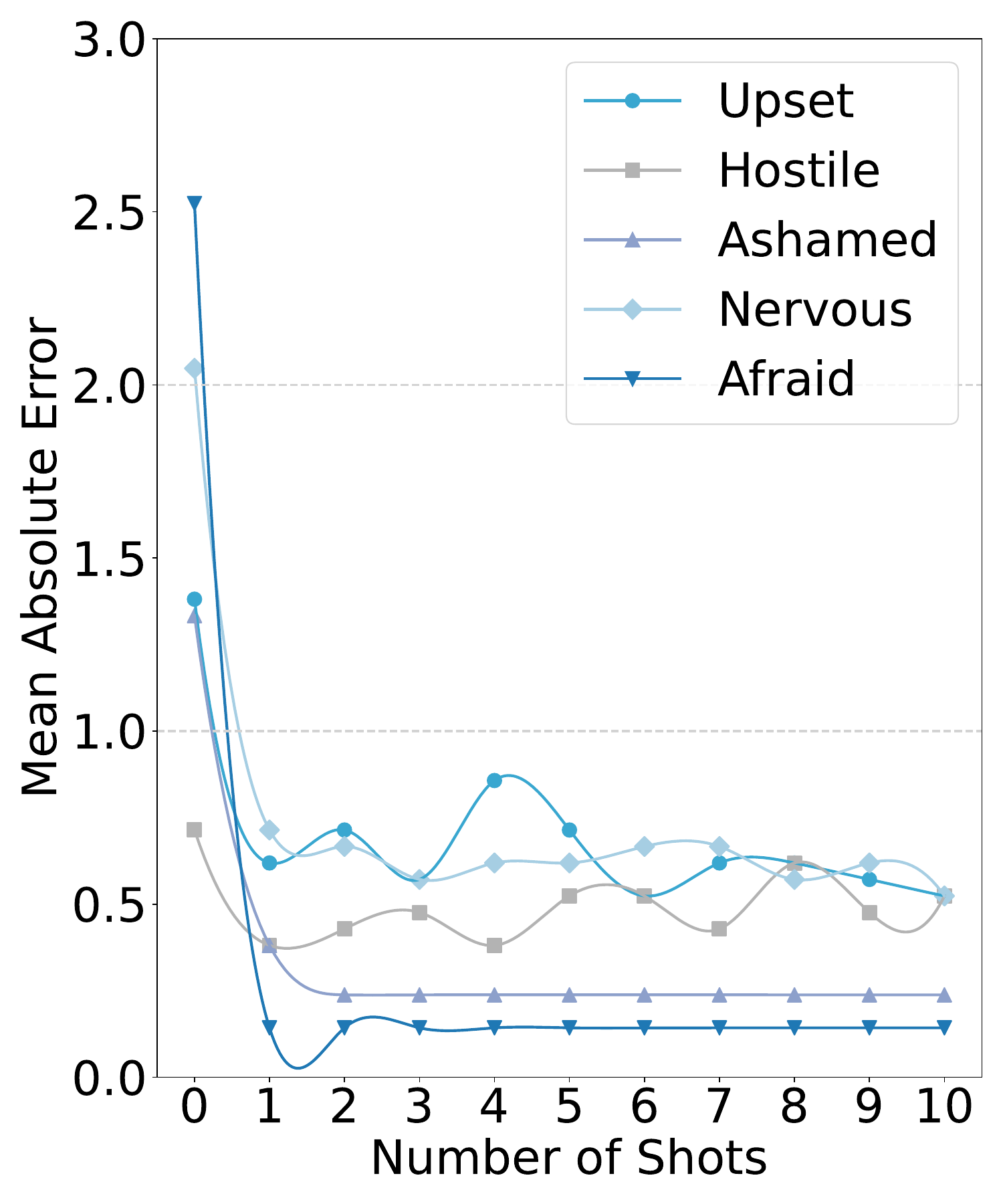}} 
  \subfigure[]{\includegraphics[trim=0pt 20pt 0pt 0pt, clip,width=0.24\textwidth]{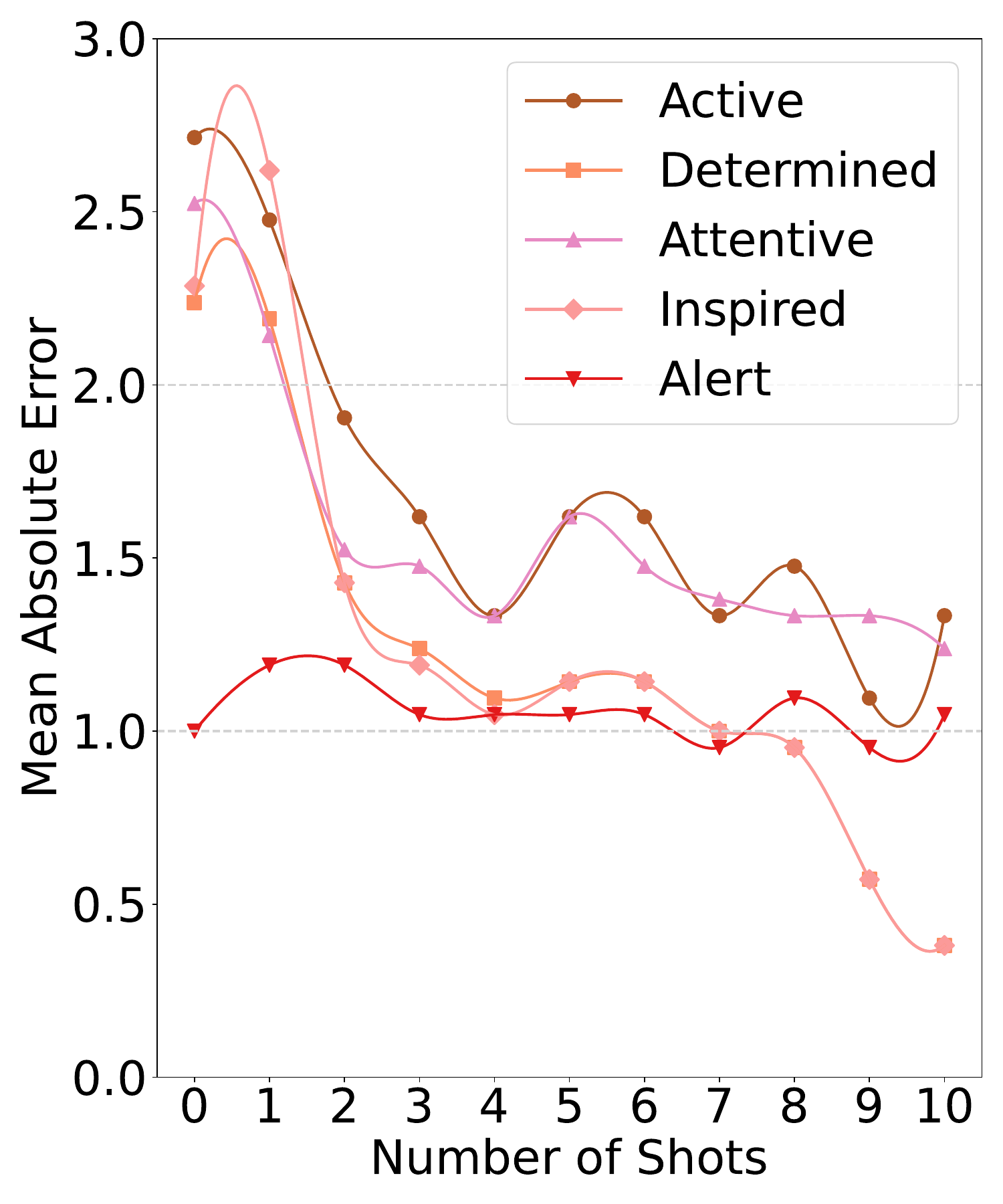}} 
  \subfigure[]{\includegraphics[trim=0pt 20pt 0pt 0pt, clip,width=0.24\textwidth]{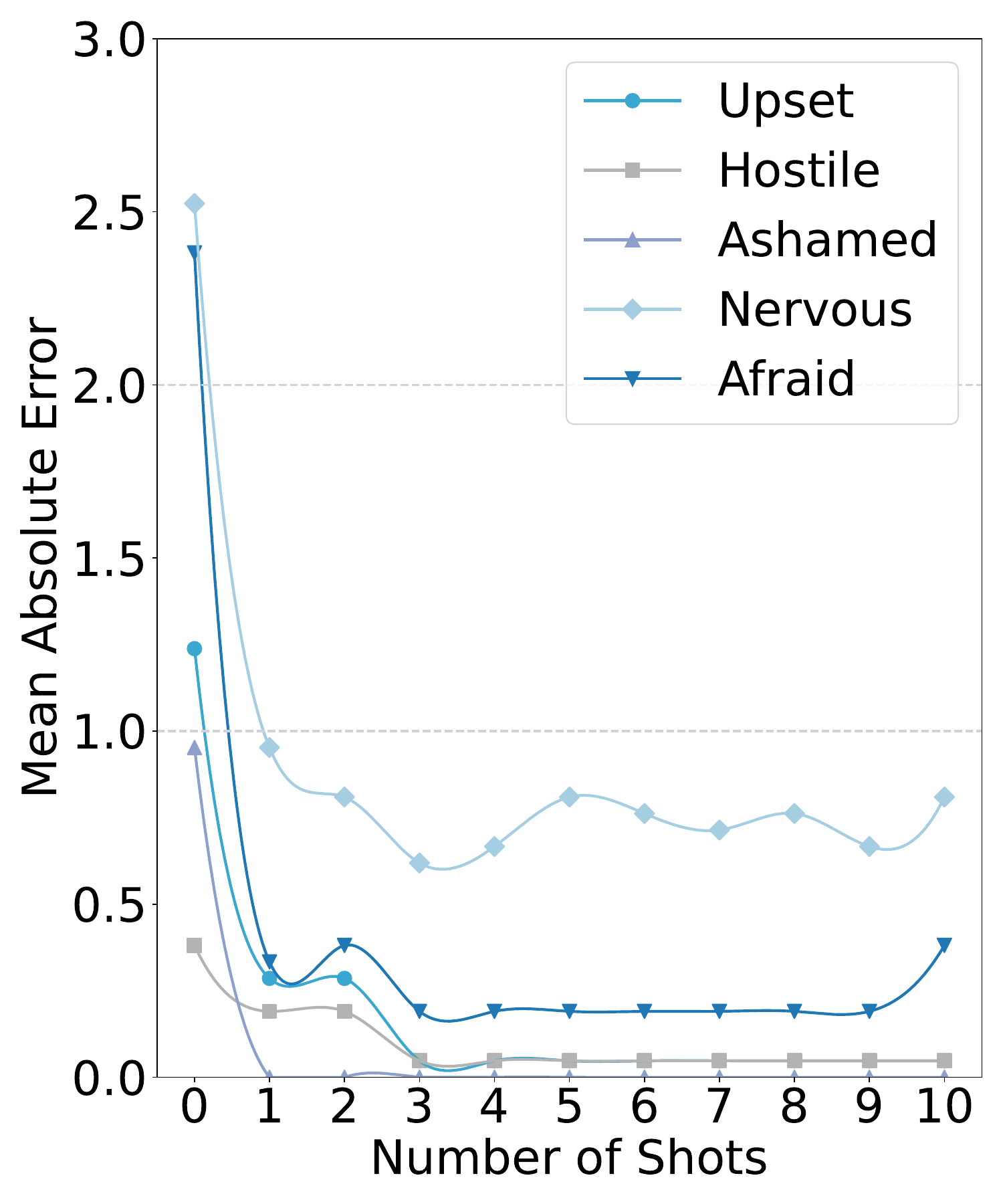}} 
  \vspace{-1.5em}
  \caption{Learning curve for data from two participants}\label{individual_examples} 
\end{figure*}

The performance improvement is also evident for each participant, with most items experiencing a decrease in MAE, although certain items prove more effective for some participants than others. For example, Figure \ref{individual_examples} displays two sets of notably diverse learning curves for the I-PANAS-SF items tested on data from two participants (a and b for one participant, c and d for the other). 

\begin{figure*}[ht]
  \centering
  \subfigure[]{\includegraphics[trim=0pt 20pt 0pt 0pt, clip,width=0.24\textwidth]{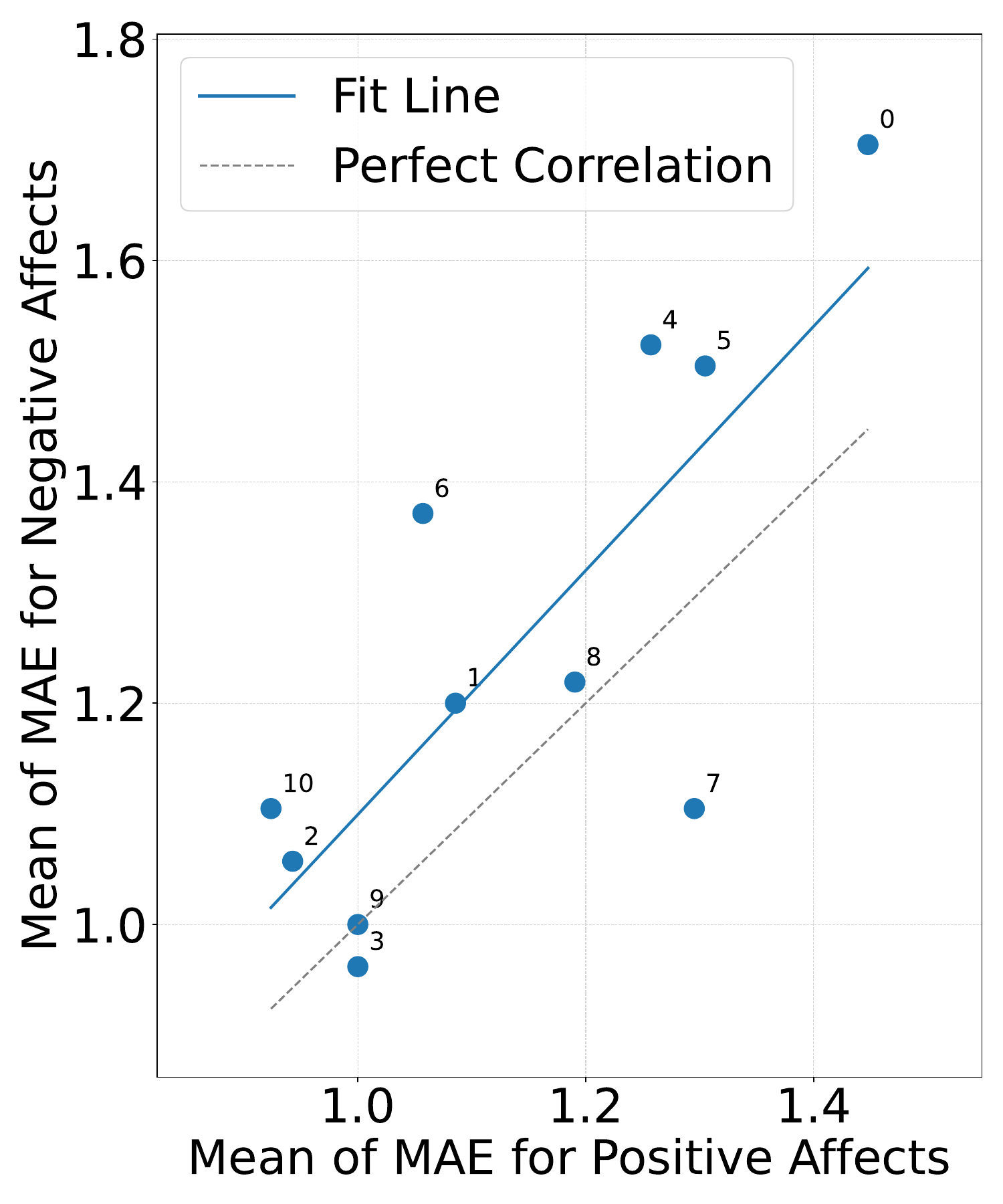}} 
  \subfigure[]{\includegraphics[trim=0pt 20pt 0pt 0pt, clip,width=0.24\textwidth]{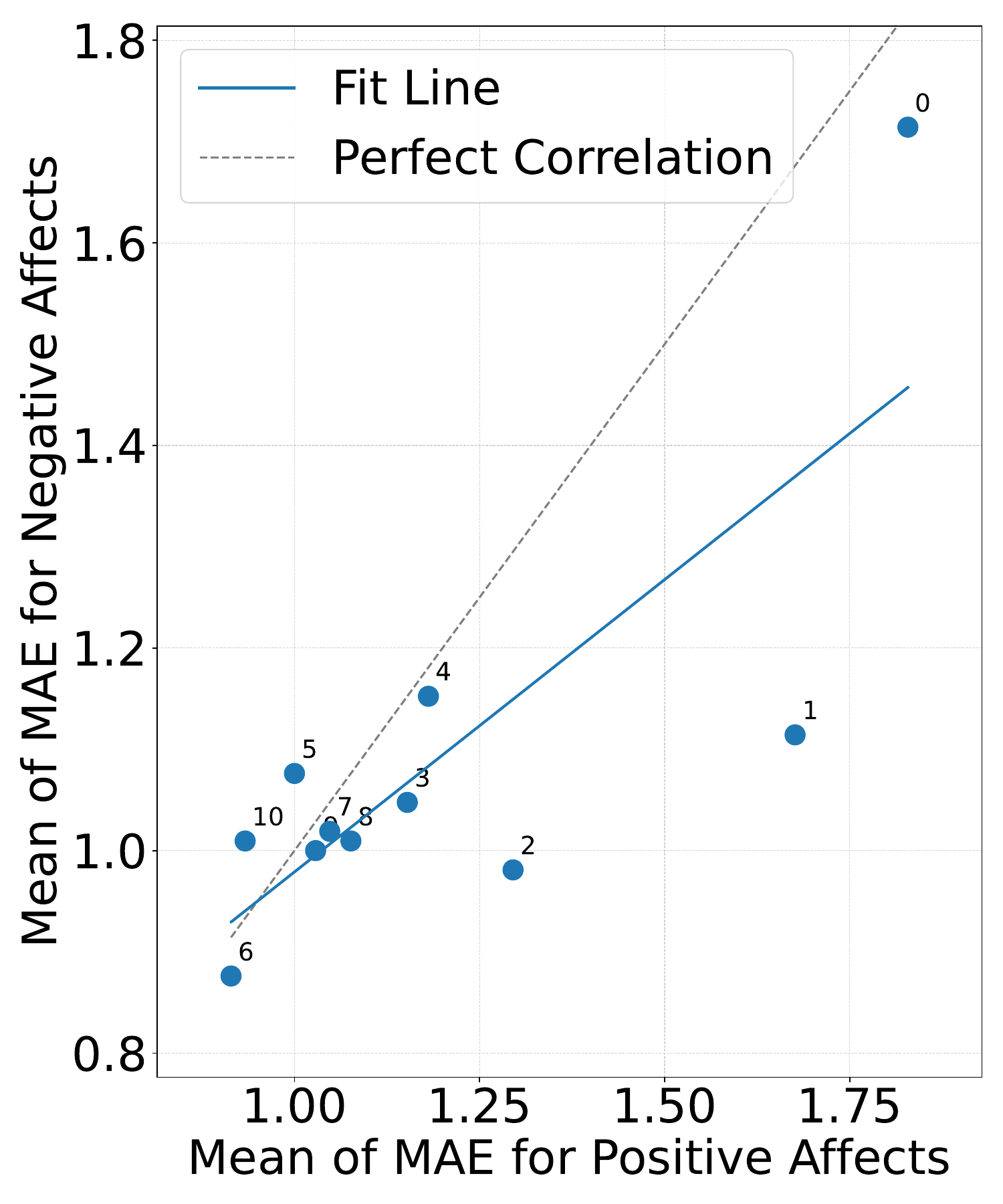}} 
  \subfigure[]{\includegraphics[trim=0pt 20pt 0pt 0pt, clip,width=0.24\textwidth]{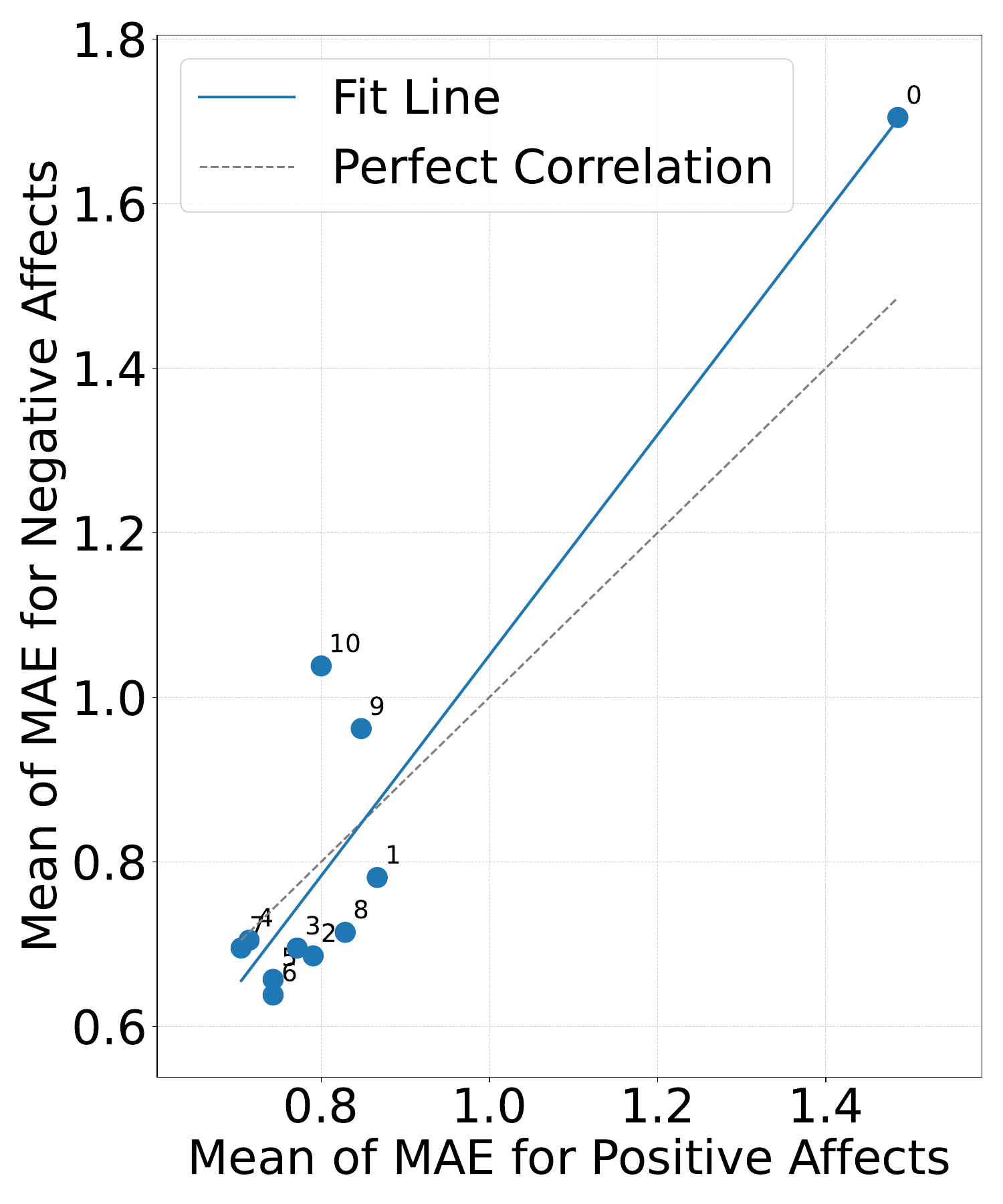}} 
  \subfigure[]{\includegraphics[trim=0pt 20pt 0pt 0pt, clip,width=0.24\textwidth]{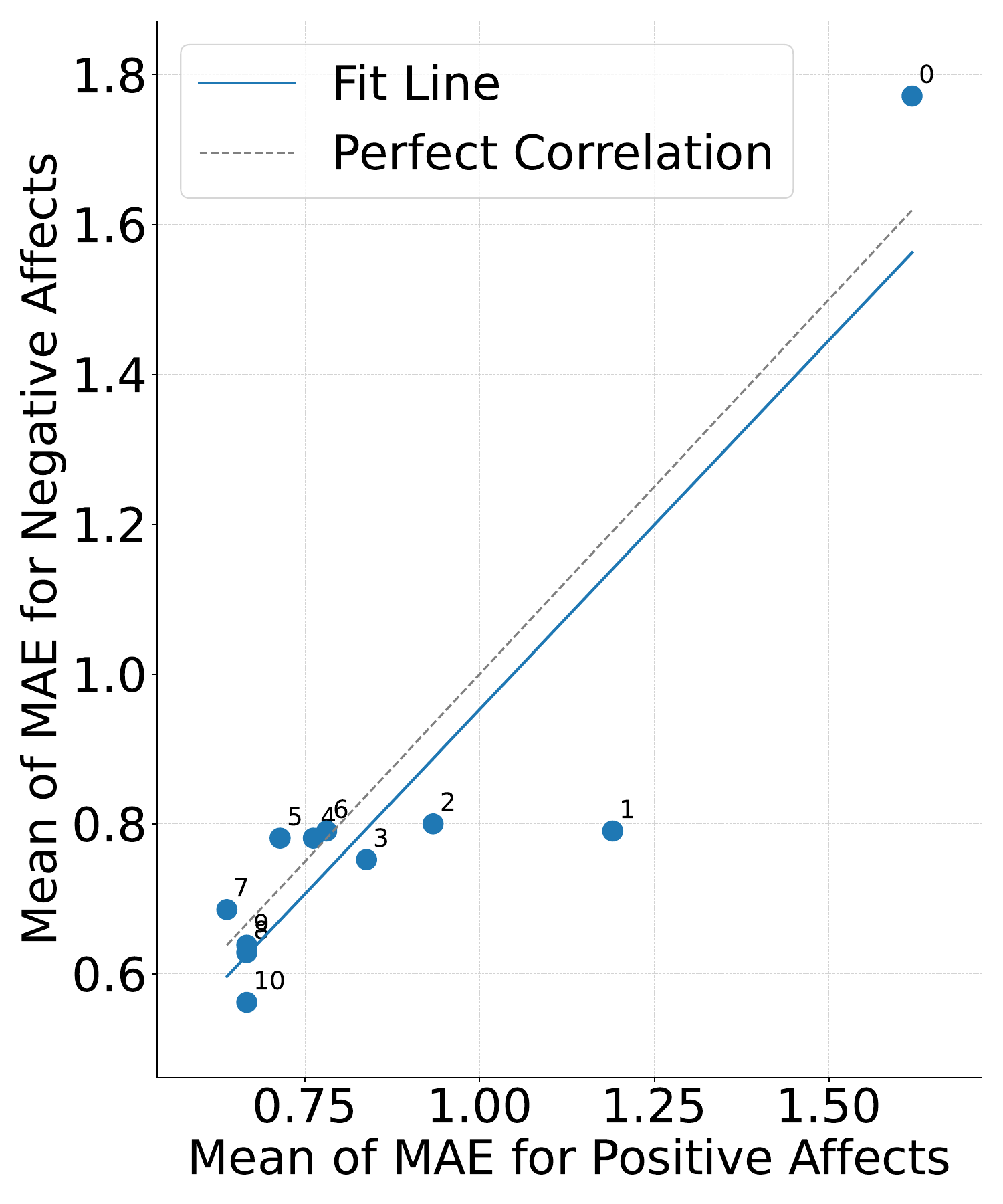}} 
  \vspace{-1.5em}
  \caption{Four participant examples of the linear relationship between MAEs for predicting positive and negative affects.}\label{pos_neg_examples} 
\end{figure*}

Furthermore, the learning curves for positive and negative affects exhibit similar trends for each individual. Figure \ref{pos_neg_examples} illustrates the correlational relationship between positive and negative affects for four participants. The performance of the LLM in predicting the affects based on smartphone behavioral features improves at a similar rate for both positive and negative affects, representing linear relationships. This consistent pattern between positive and negative affects in the learning trend indicates that each behavioral description and the corresponding affective states provided to the LLM offer similar levels of predictive information. We observe that zero-shot learning reveals larger errors (upper right corner of the graph), whereas as the number of shots increases, the points tend to shift to areas with smaller MAEs (lower left corner). This finding demonstrates that LLMs learn and perform better in prediction tasks when provided with more examples. As a result, the feasibility of using smartphone-inferred behavioral data to infer affective states with LLMs suggests a meaningful relationship between the two.


\begin{table*}[ht]
\centering
\resizebox{1\linewidth}{!} {
\begin{tabular}{ccccccccccc|cc|cc}\toprule
Shot & Active & Determined & Attentive & Inspired & Alert & Upset & Hostile & Ashamed & Nervous & Afraid & Mean & Std & Positive affects & Negative affects\\ \midrule
0 & 1.96 & 1.56 & 1.84 & 1.89 & 0.87 & 1.57 & 0.68 & 1.30 & 2.29 & 2.51 & 1.65 & 0.55 & 1.62 & 1.67\\ 
1 & 1.33 & 0.87 & 1.07 & 1.07 & 0.85 & 0.79 & 0.69 & 0.69 & 1.00 & 0.85 & 1.10 & 0.20 & 1.04 & 0.80\\ 
2 & 1.13 & 0.84 & 1.00 & 0.97 & 0.84 & 0.82 & 0.65 & 0.68 & 0.94 & 0.78 & 1.73 & 0.30 & 0.96 & 0.77\\ 
3 & 1.03 & 0.72 & 0.98 & 0.92 & 0.78 & 0.81 & 0.56 & \underline{0.61} & 0.92 & 0.77  & 1.62 & 0.28  & 0.89 & 0.73\\ 
4 & 0.97 & 0.75 & 0.99 & 0.91 & \underline{0.71}& 0.88 & 0.57 & 0.67 & 0.96 & 0.88  & 0.83 & 0.14 & 0.87 & 0.79\\ 
5 & 0.98 & 0.78 & 0.89 & 0.90 & \underline{0.71} & 0.89 & 0.60 & 0.70 & 0.92 & 0.86  & 0.82 &  0.11 & 0.85 & 0.79 \\ 
6 & 0.95 & 0.78 & 0.87 & 0.88 & 0.75 & 0.78 & 0.56 & 0.63 & 0.91 & 0.78  & 0.79 & 0.12 & 0.85 & 0.73 \\ 
7 & 0.88 & 0.80 & 0.90 & 0.84 & 0.72 & \underline{0.73} & \underline{0.53} & 0.63 & 0.94 & 0.77  & 0.79 & 0.12 & 0.83 & 0.74 \\ 
8 & 0.93 & 0.78 & 0.86 & 0.84 & 0.83 & 0.77 & 0.54 & 0.64 & 0.90 & 0.80  & 0.79 & 0.11  & 0.85 & 0.73 \\ 
9 & 0.83 & 0.68 & 0.84 & 0.83 & 0.73 & 0.73 & 0.58 & 0.66 & \underline{0.86} & 0.77  & 0.76 & \underline{0.09} & 0.79 & \underline{0.72} \\ 
10 & \underline{0.81} & \underline{0.62} & \underline{0.83} & \underline{0.76} & 0.73 & 0.80 & 0.59 & 0.64 & 0.90 & \underline{0.76} & \underline{0.75} & 0.09 & \underline{0.75} & 0.74 \\ \bottomrule
\end{tabular}}
\caption{The averaged Mean Absolute Error across the ten participants for I-PANAS-SF items. The means and standard deviations are computed across the 10 I-PANAS-SF items. The positive and negative affects are derived as averages of the MAEs for the corresponding items. The underlined results indicate the best-performing shot for each item.}\label{single_items}
\end{table*}

\textbf{Chain-of-thought.} By providing the LLM with a general understanding of the task, the model can offer conceptual insights into the relationship between individual daily activities and their affective states. Although the ultimate validity of these insights perhaps remains an open matter, they are nonetheless interesting points. Specifically, for positive affects, the model emphasizes high usage episodes and frequent screen unlocks for being active, increased app usage and long sessions with focused typing for feeling inspired, long sessions with high typing rates and specific application focus for feeling determined, and frequent typing events with short average time between keystrokes for being attentive. For negative affects, it underscores frequent screen unlocks throughout the day for alertness, an increased number of missed calls, abrupt changes in typing patterns (frequent backspacing and deletions), and increased phone calls of very short durations for feeling upset and hostile. It also highlights decreased overall phone usage and lower response rates to messages for feeling ashamed and increased numbers of very short screen unlock episodes and frequent location changes with short durations for feeling afraid.

The smartphone sensor features serve as potential indicators of individuals' affects, drawing on the LLM's prior knowledge for analysis. Leveraging these indicators, the LLM can provide logical reasoning aligned with the predicted results. The example in Figure \ref{CoT_zero} demonstrates the consistency between the predicted outcome and the logical reasoning, offering concise yet precise indications of its decisions. Among the provided sensor data (battery usage, application usage, location, calls, messages, keyboard and screen usage), the LLM's explanations emphasized involving typing events, screen unlocks, calls, and application usage. This sheds light on the perspectives guiding the LLM's analysis in predicting individuals' affects. Future endeavors could explore predictions based on more detailed or comprehensive information gathered from these sensors.

During the tasks, the LLM exhibits less confidence that the provided behavioral features could accurately uncover the negative affects compared to the positive ones. We observe a pattern of the LLM outputting lower and less varied scores (e.g., 1 or 2) for negative affects. This suggests that predicting I-PANAS-SF negative affects based on the provided features poses a challenge for the LLM.

We observed that for those providing 1 to be the predicted score, the reasoning was often "There is no clear evidence to suggest the person was \textit{<affect>} during the week". In this case, rather than assigning a random score or middle score, the LLM seems to assign a minimum value of 1 as the default value for affect items, and increases an affect item's score to 2, 3, 4 or 5 as evidence is found in the behavioral traces. Further confirmation was sought by introducing an additional option, "Not able to decide (-1)" but the LLM did not produce this answer. This suggests that the LLM possesses evidence-based analytical capabilities, producing reliable predictions in which it has the highest confidence.

Tables \ref{CoT_zero} and \ref{CoT_ten} present the results for the same participant data. As more shots are provided to the LLM, it reasons with greater evidence by comparing behavioral traces across different weeks, leading to improved performance. This demonstrates the LLM's capability to learn, summarize patterns, and adjust predictions accordingly. These findings also reveal a potential connection between smartphone behavioral features and affective states, emphasizing the importance of studying the concrete indications and associations between them.

Furthermore, in its output texts, the LLM accentuates the specifics of activities, such as the content the individual is reading on their phone, frequent application usage, and a high volume of phone unlocking episodes. It highlights that the accuracy of predictions may benefit from additional information sources, including sleep status, activity categories, specific application usage times and names, and the content of typing and viewing, rather than just speed. These features can be further explored in future studies.


\begin{table}[htbp]
\centering
\resizebox{\linewidth}{!}{%
\begin{tabular}{ccc}
\toprule
\makecell{Item \\ (True label)} & \makecell{Predicted \\ Score} &  Reasoning \\ \midrule
Active (5) & 3 & \makecell{
The person had a moderate number of unlocking \\episodes (average 8 minutes) and travelled a \\moderate distance (total travelled 4166 meters \\on 2023-08-08, 9261 meters on 2023-08-10,\\ 17230 meters on 2023-08-12) during the week.} \\ \midrule
Determined (4) & 3 & \makecell{
There is no clear evidence to suggest the person \\was determined during the week.}\\ \midrule
Attentive (5) & 2 & \makecell{
The person did not use any social media \\ applications or dating applications, and there\\ were no emails or messages sent or received \\ during this period. This suggests they were not \\ attentive to these aspects of their phone.} \\ \midrule
Inspired (4) & 1 & \makecell{
There is no clear evidence to suggest the person \\ was inspired during the week.} \\ \midrule
Alert (4) & 3 & \makecell{
The person had a moderate number of unlocking \\ episodes (average 6 minutes across the week) and\\ travelled a moderate distance (total \\travelled ranging from 4166 meters on \\2023-08-08 to 17230 meters on 2023-08-12) which could \\indicate some alertness.}\\\midrule
Upset (3) & 1 & \makecell{
There is no clear evidence to suggest the person \\was upset during the week.}\\\midrule
Hostile (2) & 1 & \makecell{
There is no clear evidence to suggest the person\\ was hostile during the week.}\\\midrule
Ashamed (1) & 1 & \makecell{
There is no clear evidence to suggest the person \\was ashamed during the week.}\\\midrule
Nervous (5) & 2 & \makecell{
There is no clear evidence to suggest the person \\was nervous during the week. They did not use \\social media or dating applications, and \\there were no emails or messages sent or received.}\\\midrule
Afraid (4) & 1 & \makecell{
There is no clear evidence to suggest the person \\was afraid during the week.}\\ \bottomrule
\end{tabular}
}
\caption{Zero-Shot Chain-of-Thought reasoning for one participant.}\label{CoT_zero}
\end{table}

\begin{table}[htbp]
\centering
\resizebox{\linewidth}{!}{%
\begin{tabular}{ccc}
\toprule
\makecell{Item \\ (True label)} & \makecell{Predicted \\ Score}  & Reasoning \\ \midrule
Active (5) & 4 & \makecell{The student's average time spent moving between locations\\ is consistently around 10 km/h, and they have a significant \\amount of total traveled meters (often over 10,000 meters). \\These indicate a generally active lifestyle. While their\\ future week involves slightly less movement than usual, \\it is still above average, so I predict a score of 4.
} \\ \midrule
Determined (4) & 4 & \makecell{
The student frequently uses applications for extended periods.\\ This suggests a consistent engagement with tasks, potentially \\indicating a sense of determination. The future week involves \\similar levels of application usage, making a score of 4 \\appropriate. 
}\\ \midrule
Attentive (5) & 4 & \makecell{
The student's average time spent at the most visited location is \\very high, often exceeding 1000 minutes. This could imply a \\focus on tasks or activities within this location. This pattern\\ persists in the future week. Thus, a score of 4 is likely. 
} \\ \midrule
Inspired (4) & 4 & \makecell{
While there isn't direct evidence in the provided data, the \\student's frequent use of applications, particularly for extended \\periods, suggests a potential engagement with learning or \\creative pursuits. The future week maintains this trend, so a \\score of 4 seems fitting. 
} \\ \midrule
Alert (4) & 3 & \makecell{
The student has a relatively low number of unlock episodes per day\\ and a low average duration of unlock episodes. This could \\indicate a less frequent need to access the phone, suggesting a \\possible lower level of alertness. The future week maintains \\this pattern, so I predict a score of 3. 
}\\\midrule
Upset (3) & 3 & \makecell{
The student often experiences missed calls, suggesting potential \\disruptions and stress. In the future week, there is a small \\increase in missed calls, so I predict a score of 3.
}\\\midrule
Hostile (2) & 3 & \makecell{
There is no clear indication of hostile behaviors based on the \\provided data. The student has consistent low numbers of missed\\ and outgoing calls, suggesting a general calmness. However, the \\future week sees a small uptick in missed calls, potentially \\signaling a slight increase in tension, making a score of 3 \\more likely. 
}\\\midrule
Ashamed (1) & 2 & \makecell{
No clear indicators of shame can be found in the provided data. \\The student's low levels of social media and dating app usage \\suggest a possible preference for privacy or a lack of social \\engagement. I predict a score of 2 for the future week. 
}\\\midrule
Nervous (5) & 4 & \makecell{
The student often shows a high number of typing events per session. \\This could indicate a possible nervousness or anxiousness, \\particularly if there are many events where the length of the \\text changes in fewer characters, suggesting possible backspacing\\ or edits. The future week reflects this pattern, so a score \\of 4 seems appropriate.
}\\\midrule
Afraid (4) & 3 & \makecell{
The provided data lacks clear indicators of fear. The student's \\consistent low numbers of missed and outgoing calls suggest a \\general feeling of safety and security. However, the future \\week features a slight increase in missed calls, which might \\indicate a potential for feeling slightly more vulnerable or \\anxious, leading to a score of 3.
}\\ \bottomrule
\end{tabular}
}
\caption{Ten-Shot Chain-of-Thought reasoning for one participant}\label{CoT_ten}
\end{table}

\vspace{-3em}
\section{Future Work and Conclusions}
In this study, we demonstrated the potential of LLMs in harnessing smartphone-collected behavioral data to predict affective states of university students by leveraging both zero-shot and few-shot learning approaches. Our results indicate that the LLM's ability to perform zero-shot tasks for mental wellbeing based on weekly summarized behaviors suffers from a relatively high error rate. However, when provided with labeled information, the LLMs achieve significantly improved predictions. This positions LLMs as an invaluable resource for future research and practical applications in wellbeing and other fields, deepening our understanding of individuals' mental health, emotions, and affective states.

For future work, fine-tuning tasks could be conducted to develop a model driven by daily activities, thereby incorporating more data and possibly enabling a comparison between individual and general predictive models. Given the subjective nature of self-reported datasets, models may exhibit biases stemming from imbalanced class distributions, with some classes being significantly underrepresented compared to others. Hence, increasing the dataset size or employing resampling techniques becomes imperative to construct a more robust model overall. Furthermore, leveraging LLMs to predict individual items of additional psychometric measures for prediction tasks would be advisable, facilitating the construction of more nuanced insights into people's affective states at a wellbeing level.


\vspace{3em}
\section{Appendix}
\begin{table}[htbp]
\centering
\resizebox{\linewidth}{!}{%
\begin{tabular}{ccc}
\toprule
        Item & Feature & Sensor \\ \midrule
        1 & the count of episodes using Email applications & \multirow{20}{*}{Applications} \\ \cmidrule{1-2}
        2 & the count of episodes using all applications &  \\ \cmidrule{1-2}
        3 & the count of episodes using social media applications &  \\ \cmidrule{1-2}
        4 & the count of episodes using dating applications &  \\ \cmidrule{1-2}
        5 & the count of episodes using social applications &  \\ \cmidrule{1-2}
        6 & the count of episodes using entertainment applications &  \\ \cmidrule{1-2}
        7 & the count of episodes using Facebook Moments &  \\ \cmidrule{1-2}
        8 & the count of usage episodes for the application with top usage &  \\ \cmidrule{1-2}
        9 & the count of episodes using YouTube &  \\ \cmidrule{1-2}
        10 & the count of episodes using Twitter &  \\ \cmidrule{1-2}
        11 & the duration of using Email &  \\ \cmidrule{1-2}
        12 & the duration of using all applications &  \\ \cmidrule{1-2}
        13 & the duration using social media applications &  \\ \cmidrule{1-2}
        14 & the duration of using dating applications &  \\ \cmidrule{1-2}
        15 & the duration of using social applications &  \\ \cmidrule{1-2}
        16 & the duration of using entertainment applications &  \\ \cmidrule{1-2}
        17 & the duration of using Facebook Moments &  \\ \cmidrule{1-2}
        18 & the duration of using the application with top usage &  \\ \cmidrule{1-2}
        19 & the duration of using YouTube &  \\ \cmidrule{1-2}
        20 & the duration of using Twitter &  \\ \midrule

        21 & the count of discharging episodes & \multirow{2}{*}{Battery} \\ \cmidrule{1-2}
        22 & the count of charging episodes &  \\ \midrule
        
        23 & the number of missed calls & \multirow{15}{*}{Calls} \\ \cmidrule{1-2}
        24 & the number of distinct contacts associated with missed calls &  \\ \cmidrule{1-2}
        25 & the number of missed calls from the most frequent contact &  \\ \cmidrule{1-2}
        26 & the number of incoming calls &  \\ \cmidrule{1-2}
        27 & the number of distinct contacts associated with incoming calls &  \\ \cmidrule{1-2}
        28 & the mean of incoming call duration &  \\ \cmidrule{1-2}
        29 & the sum of incoming call duration &  \\ \cmidrule{1-2}
        30 & the mode of incoming call duration &  \\ \cmidrule{1-2}
        31 & the count of incoming calls from the most frequent contact &  \\ \cmidrule{1-2}
        32 & the number of outgoing calls &  \\ \cmidrule{1-2}
        33 & the number of distinct contacts associated with outgoing calls &  \\ \cmidrule{1-2}
        34 & the mean of outgoing call duration &  \\ \cmidrule{1-2}\cmidrule{1-2}
        35 & the sum of outgoing call duration &  \\ \cmidrule{1-2}
        36 & the mode of outgoing call duration &  \\ \cmidrule{1-2}
        37 & the count of outgoing calls from the most frequent contact &  \\ \midrule

        38 & the count of typing events & \multirow{7}{*}{Keyboard} \\ \cmidrule{1-2}
        39 & \makecell{the count of keyboard typing or swiping event\\ where the length changes exactly one more character} &  \\ \cmidrule{1-2}
        40 & \makecell{the count of keyboard typing or swiping event where\\ the length changes less than one fewer character} &  \\ \cmidrule{1-2}
        41 & \makecell{the count of keyboard typing or swiping event where\\ the length changes exactly one fewer character} &  \\ \cmidrule{1-2}
        42 & the number of characters in average in a session &  \\ \cmidrule{1-2}
        43 & the number of typing sessions &  \\ \cmidrule{1-2}
        44 & the average time between keystrokes &  \\ \bottomrule
\end{tabular}
}
\caption{Smartphone sensor features embedded in the prompts.}\label{sensor_features}
\end{table}

\begin{table}[htbp]
\centering
\resizebox{\linewidth}{!}{%
\begin{tabular}{ccc}
\toprule
        45 & time spent at the second most visited location & \multirow{20}{*}{Locations} \\ \cmidrule{1-2}
        46 & maximum time spent at any location cluster &  \\ \cmidrule{1-2}
        47 & \makecell{the ratio of time spent moving between\\ locations to time spent stationary at a location} &  \\ \cmidrule{1-2}
        48 & total travelled distance &  \\ \cmidrule{1-2}
        49 & standard deviation of the time spent at location clusters &  \\ \cmidrule{1-2}
        50 & time spent at the most visited location &  \\ \cmidrule{1-2}
        51 & average time spent at location clusters &  \\ \cmidrule{1-2}
        52 & normalized entropy of location visits &  \\ \cmidrule{1-2}
        53 & variance of speed during movement between locations &  \\\cmidrule{1-2} 
        54 & time spent at home &  \\ \cmidrule{1-2}
        55 & time spent at the third most visited location &  \\ \cmidrule{1-2}
        56 & the number of transitions between distinct locations &  \\ \cmidrule{1-2}
        57 & minimum time spent at any location cluster &  \\ \cmidrule{1-2}
        58 & radius of Gyration (RoG) indicating the area covered &  \\ \cmidrule{1-2}
        59 & average speed during movement between locations &  \\ \cmidrule{1-2}
        60 & percent of time considered outliers in location data &  \\ \cmidrule{1-2}
        61 & the variance of locations visited &  \\ \cmidrule{1-2}
        62 & the logarithm of the variance of locations visited &  \\ \cmidrule{1-2}
        63 & the number of significant places visited &  \\ \cmidrule{1-2}
        64 & the entropy of location visits &  \\ \midrule

        65 & the number of received messages from the most frequent contact & \multirow{6}{*}{Messages} \\ \cmidrule{1-2}
        66 & the number of received messages &  \\ \cmidrule{1-2}
        67 & the number of distinct contacts associated with received messages &  \\ \cmidrule{1-2}
        68 & the number of sent messages to the most frequent contact &  \\ \cmidrule{1-2}
        69 & the number of sent messages &  \\ \cmidrule{1-2}
        70 & the number of distinct contacts associated with sent messages &  \\ \midrule

        71 & the number of unlock episodes & \multirow{7}{*}{Screen status} \\ \cmidrule{1-2}
        72 & total duration of unlock episodes &  \\ \cmidrule{1-2}
        73 & the length of longest unlock episode &  \\ \cmidrule{1-2}
        74 & average time of unlock episodes &  \\ \cmidrule{1-2}
        75 & the length of shortest unlock episode &  \\ \cmidrule{1-2}
        76 & standard deviation of unlock episodes &  \\ \cmidrule{1-2}
        77 & time between the first unlock episode and midnight &  \\ \bottomrule
\end{tabular}
}
\caption{Smartphone sensor features embedded in the prompts Cont.}\label{sensor_features2}
\end{table}

\newpage
\bibliographystyle{ACM-Reference-Format}
\bibliography{Reference.bib}

\end{document}